\documentclass[aps,preprint,tightenlines,showpacs,nofootinbib]{revtex4}
\usepackage{epsfig}
\usepackage{multirow}
\usepackage{amsmath,stackrel}
\usepackage{amssymb}
\begin{document}
\title{Multiquark contributions to charm baryon spectroscopy}
\author{T.~F.~Caram\'es}
\affiliation{Departamento de F{\'\i}sica Fundamental,
Universidad de Salamanca, 37008 Salamanca, Spain}
\author{A.~Valcarce}
\affiliation{Departamento de F{\'\i}sica Fundamental,
Universidad de Salamanca, 37008 Salamanca, Spain}
\date{\emph{Version of }\today}

\begin{abstract}
We study possible multiquark contributions to the charm baryon spectrum 
by considering higher order Fock space components. 
For this purpose we perform two different calculations. In a first approach
we do a coupled-channel calculation of the $ND$ system looking for molecular states. 
In a second step we allow for the coupling to a heavy baryon--light meson two-hadron system
looking for compact exotic five--quark structures. Both calculations have been done
within the framework of a chiral constituent quark model.
The model, tuned in the description of the baryon and meson spectra 
as well as the $NN$ interaction, provides parameter-free predictions for 
charm $+1$ molecular or compact two-hadron systems. 
Unlike the $N \bar D$ system, no sharp quark-Pauli effects are found.
However, the existence of different two-hadron thresholds for the five-quark
system will make the coupled-channel dynamics relevant. 
Only a few channels are candidates to lodge molecular or compact hadrons with
a five-quark structure, being specially relevant the $(T)J^P=(0)1/2^-$ and
$(T)J^P=(2)5/2^-$ channels.
The identification of molecular states and/or compact hadrons with multiquark 
components either with or without exotic quantum numbers is a challenge
of different collaborations like $\bar {\rm P}$ANDA, LHCb, ExHIC or J-PARC.
\end{abstract}

\maketitle

\section{Introduction}
\label{secI}

One of the most basic problems of QCD is to identify all the clusters of quarks,
antiquarks and gluons that are sufficiently bound by QCD interactions that they
are either stable particles or appropriately long-lived to be observed as resonances~\cite{Bra13}.
To this respect, charm hadron physics has become a cornerstone due to the 
experimental findings during the last decade. On the one hand one encounters
the outstanding discovery in charmonium spectroscopy 
of the flagship of the so-called $XYZ$ states, the $X(3872)$~\cite{Cho03}. 
Before this discovery, and based on Gell-Mann conjecture~\cite{Gel64}, the hadronic experimental 
data were classified either as $q\overline{q}$ or $qqq$ states according to $SU(3)$ irreducible 
representations. However, since 2003 more than twenty meson resonances reported 
by different experimental collaborations, most of them close to a two-meson threshold,
had properties that made a simple quark-antiquark structure unlikely~\cite{Bod13}. Although
this observation could be coincidental due to the large number of thresholds in the energy region
where the $XYZ$ mesons have been reported, it could also point to
a close relation between some particular thresholds and resonances contributing to the standard
quark-antiquark heavy meson spectroscopy. On the other hand, 
a similar situation has arisen in the charm baryon spectrum during the last years with the advent of a large set
of experimental data (see Refs.~\cite{Kle10,Cre13} for a comprehensive update 
of the experimental and theoretical situation of the heavy baryon spectra). The properties of some
excited states show an elusive nature as three-quark systems. Likewise the
charmonium spectrum, some of them are rather close to a baryon-meson threshold suggesting a 
possible molecular or compact structure~\cite{Hel07,Hel10,Don10,Yam13,Alb13,Zha14,ChaYY}.
It has been already highlighted within a simple toy model the key role 
that $S$ wave meson-baryon thresholds may play in matching 
poor light-baryon mass predictions from quark models with data~\cite{Gon08}.
Thus, the analysis of possible multiquark contributions close to meson-baryon 
thresholds with a full-fledged quark dynamical model could help in the 
understanding of heavy baryon spectroscopy. 

The existence of molecular contributions in the charm baryon spectrum stem primarily on
the interaction between charm mesons and nucleons, what on the other hand has turned into an interesting
subject in several contexts~\cite{Pan09}. It is particularly interesting for the study
of chiral symmetry restoration in a hot and/or dense medium~\cite{Kog83}. It will also help in the
understanding of the suppression of the $J/\Psi$ production in heavy ion collisions~\cite{Mat86}. Besides,
it may shed light on the possible existence of exotic nuclei with heavy flavors~\cite{Dov77,Yok14}. Experimentally, 
it will become possible to analyze the interaction of charm mesons with nucleons inside 
nuclear matter with the operation of the FAIR facility at the GSI laboratory in Germany~\cite{Pan09}. 
There are proposals for experiments by the $\bar {\rm P}$ANDA Collaboration to produce $D$ mesons by 
annihilating antiprotons on the deuteron. This could be achieved with an antiproton beam, by tuning 
the antiproton energy to one of the higher-mass 
charmonium states that decays into open charm mesons. These experimental ideas may become plausible based on 
recent estimations of the cross section for the production of $D\bar D$ pairs in proton-antiproton
collisions~\cite{Kho12}. There are also different theoretical estimations about the
production rate at $\bar {\rm P}$ANDA of $\Lambda_c$ baryons through the direct process
$p\bar p \to \Lambda_c \bar\Lambda_c$~\cite{Heo11,Hai11}. The identification of hadronic molecular states and/or hadrons with multiquark 
components either with or without exotic quantum numbers is also a challenge
in relativistic heavy ion collisions offering a promising resolution to this 
problem~\cite{Cho11,Cho11b}. Besides the LHCb~\cite{Mil13} and CDF~\cite{Aal14} Collaborations 
are providing a huge dataset of new measurements of heavy flavor spectroscopy.
In the coming future, J-PARC also intends to contribute to the experimental
measurement of exotic baryons. Thus, a good knowledge of the interaction of charm mesons 
with ordinary hadrons, like nucleons or $\Delta's$, is a prerequisite. 

Before one can infer in a sensitive manner changes of the interaction in the medium~\cite{Sai07,Car14}, a reasonable 
understanding of the interaction in free space is required. However, here one has to manage
with an important difficulty, namely the complete lack of experimental data at low energies for the
free-space interaction. Thus, the generalization of models describing the two-hadron interaction
in the light flavor sector could offer insight about the unknown interaction of hadrons with
heavy flavors. This is the main purpose of this work, to make use of a chiral constituent quark
model describing the $NN$ interaction~\cite{Val05} as well as the meson spectrum in all 
flavor sectors~\cite{Vij05} to obtain parameter-free predictions that may be testable in 
future experiments. Such a project was already
undertaken for the interaction between two charm mesons~\cite{Car09} and also for the 
interaction between anticharm
mesons and nucleons~\cite{Car12} what encourages us in the present challenge.

The paper is organized as follows. In Sec.~\ref{secII} we will first present a brief description
of the quark-model wave function for the baryon-meson system. We will later on 
revisit the interacting potential and finally we will summarize
the solution of the two-body problem by means of the Fredholm determinant.
In Sec.~\ref{secV} we present and discuss our results. We will first briefly discuss the baryon-meson 
interaction in comparison to hadronic models. We will analyze the character 
of the different isospin-spin channels, looking for the attractive 
ones that may lodge resonances either as a molecule or as a compact five--quark
state, to be measured by experiment.
We will also compare with existing results in the literature.
Finally, in Sec.~\ref{secVI} we summarize our main conclusions.

\section{The baryon-meson system}
\label{secII}
\subsection{The baryon-meson wave function}
\label{secIIA}

In order to describe the baryon-meson system we shall use a constituent quark
cluster model, i.e., hadrons are described as clusters of
quarks and antiquarks. Assuming a two-center shell model the wave function of
an arbitrary baryon-meson system, a baryon $B_i$ and a meson $M_j$, can be 
written as:
\begin{equation}
\Psi_{B_i M_j}^{LST}({\vec R}) =  {\cal A}
\left[ B_i \left( 123;{-{\frac{{\vec R} 
}{2}}} \right) M_j \left( 4\bar{5}; {+\frac{{\vec R} }{2}} \right) \right]^{LST}
\, ,  \label{Gor}
\end{equation}
where ${\cal A}$ is the antisymmetrization operator accounting for 
the possible existence of identical quarks inside the hadrons. In the case we are
interested in, baryon-meson systems made of $N$ or $\Delta$ baryons and 
$D$ or $D^*$ mesons, no identical quarks can be exchanged between the baryon and the
meson and thus no sharp quark-Pauli effects are expected.

If we assume gaussian $0s$ wave functions for the quarks inside the hadrons,
the normalization of the baryon-meson wave function $\Psi_{B_i M_j}^{LST}({\vec R})$
of Eq.~(\ref{Gor}) can be expressed as,

\begin{equation}
{\cal N}_{B_iM_j}^{LST}(R) = 4 \pi \exp \left\lbrace {-\frac{R^2}{8} 
\left( \frac{4}{b^2} + \frac{1}{b_c^2}\right)}\right\rbrace  i_{L+1/2} 
\left[ \frac{R^2}{8} \left( \frac{4}{b^2} + \frac{1}{b_c^2} \right)\right] \, , \\
\label{Norm2}
\end{equation}
where, for the sake of generality, we have assumed different gaussian parameters
for the wave function of the light quarks ($b$) and the heavy quark ($b_c$).
In the limit where the two hadrons overlap ($R \to 0$), the Pauli principle
does not impose any antisymmetry requirement. This can be easily checked
for the $L=0$ partial waves, where such effects would be prominent. Using the asymptotic form
of the Bessel functions, $i_{L+1/2}$, we obtain the $S$ wave normalization kernel in the overlapping 
region that behaves like a constant for $R=0$,
\begin{equation}
{\cal N}_{B_iM_j}^{L=0ST} \stackrel[R\to 0]{}{\hbox to 20pt{\rightarrowfill}} 4\pi \left\lbrace {1 -\frac{R^2}{8} 
\left( \frac{4}{b^2} + \frac{1}{b_c^2}\right)}\right\rbrace  
\left\lbrace 1  +  \frac{1}{6} \left( \frac{R^2}{8 b_c^2}\right)^2  
\left( 1+\frac{4b_c^2}{b^2}\right)^2  + ... \right\rbrace \,.
\end{equation}

\subsection{The two-body interactions}
\label{secIII}

The two-body interactions involved in the study of the baryon-meson
system are obtained from the chiral constituent quark model~\cite{Val05}. 
This model was proposed in the early 90's in an attempt to
obtain a simultaneous description of the nucleon-nucleon
interaction and the baryon spectra. It was later on generalized to all 
flavor sectors ~\cite{Vij05}. 
In this model hadrons are described as clusters of three interacting 
massive (constituent) quarks, the mass coming from the spontaneous breaking 
of the original $SU(2)_{L}\otimes SU(2)_{R}$ 
chiral symmetry of the QCD Lagrangian.
QCD perturbative effects are taken into account
through the one-gluon-exchange (OGE) potential~\cite{Ruj75}.
It reads,  
\begin{equation}
V_{\rm OGE}({\vec{r}}_{ij})=
        {\frac{\alpha_s}{4}}\,{\vec{\lambda}}_{i}^{\rm
c} \cdot {\vec{\lambda}}_{j}^{\rm c}
        \Biggl \lbrace{ \frac{1}{r_{ij}}}
        - \dfrac{1} {4} \left(
{\frac{1}{{2\,m_{i}^{2}}}}\, +
{\frac{1}{{2\,m_{j}^{2}}}}\,
        + {\frac{2 \vec \sigma_i \cdot \vec
\sigma_j}{3 m_i m_j}} \right)\,\,
          {\frac{{e^{-r_{ij}/r_{0}}}}
{{r_{0}^{2}\,\,r_{ij}}}}
        - \dfrac{3 S_{ij}}{4 m_q^2 r_{ij}^3}
        \Biggr \rbrace\,\, ,
\end{equation}
where $\lambda^{c}$ are the $SU(3)$ color matrices, 
$r_0=\hat r_0/\mu$ is a flavor-dependent regularization scaling with the 
reduced mass of the interacting pair, and $\alpha_s$ is the
scale-dependent strong coupling constant given by~\cite{Vij05},
\begin{equation}
\alpha_s(\mu)={\frac{\alpha_0}{\rm{ln}\left[{({\mu^2+\mu^2_0})/
\gamma_0^2}\right]}},
\label{asf}
\end{equation}
where $\alpha_0=2.118$, 
$\mu_0=36.976$ MeV and $\gamma_0=0.113$ fm$^{-1}$. This equation 
gives rise to $\alpha_s\sim0.54$ for the light-quark sector
and $\alpha_s\sim0.43$ for $uc$ pairs.

Non-perturbative effects are due to the spontaneous breaking of the original 
chiral symmetry at some momentum scale. In this domain of momenta, light quarks 
interact through Goldstone boson exchange potentials,
\begin{equation}
V_{\chi}(\vec{r}_{ij})\, = \, V_{\rm OSE}(\vec{r}_{ij}) \, + \, V_{\rm OPE}(\vec{r}_{ij}) \, ,
\end{equation}
where
\begin{eqnarray}
V_{\rm OSE}(\vec{r}_{ij}) &=&
    -\dfrac{g^2_{\rm ch}}{{4 \pi}} \,
     \dfrac{\Lambda^2}{\Lambda^2 - m_{\sigma}^2}
     \, m_{\sigma} \, \left[ Y (m_{\sigma} \,
r_{ij})-
     \dfrac{\Lambda}{{m_{\sigma}}} \,
     Y (\Lambda \, r_{ij}) \right] \,, \nonumber \\
V_{\rm OPE}(\vec{r}_{ij})&=&
     \dfrac{ g_{\rm ch}^2}{4
\pi}\dfrac{m_{\pi}^2}{12 m_i m_j}
     \dfrac{\Lambda^2}{\Lambda^2 - m_{\pi}^2}
m_{\pi}
     \Biggr\{\left[ Y(m_{\pi} \,r_{ij})
     -\dfrac{\Lambda^3}{m_{\pi}^3} Y(\Lambda
\,r_{ij})\right]
     \vec{\sigma}_i \cdot \vec{\sigma}_j 
\nonumber \\
&&   \qquad\qquad +\left[H (m_{\pi} \,r_{ij})
     -\dfrac{\Lambda^3}{m_{\pi}^3} H(\Lambda
\,r_{ij}) \right] S_{ij}
     \Biggr\}  (\vec{\tau}_i \cdot \vec{\tau}_j)
\, .
\end{eqnarray}
$g^2_{\rm ch}/4\pi$ is the chiral coupling constant,
$Y(x)$ is the standard Yukawa function defined by $Y(x)=e^{-x}/x$,
$S_{ij} \, = \, 3 \, ({\vec \sigma}_i \cdot
{\hat r}_{ij}) ({\vec \sigma}_j \cdot  {\hat r}_{ij})
\, - \, {\vec \sigma}_i \cdot {\vec \sigma}_j$ is
the quark tensor operator, and $H(x)=(1+3/x+3/x^2)\,Y(x)$.

Finally, any model imitating QCD should incorporate
confinement. Being a basic term from the spectroscopic point of view
it is negligible for the hadron-hadron interaction. Lattice calculations 
suggest a screening effect on the potential when increasing the interquark 
distance~\cite{Bal01},
\begin{equation}
V_{\rm CON}(\vec{r}_{ij})=\{-a_{c}\,(1-e^{-\mu_c\,r_{ij}})\}(\vec{%
\lambda^c}_{i}\cdot \vec{ \lambda^c}_{j})\, .
\end{equation}
Once perturbative (one-gluon exchange) and nonperturbative (confinement and
chiral symmetry breaking) aspects of QCD have been considered, one ends up with
a quark-quark interaction of the form 
\begin{equation} 
V_{q_iq_j}(\vec{r}_{ij})=\left\{ \begin{array}{ll} 
\left[ q_iq_j=nn \right] \Rightarrow V_{\rm CON}(\vec{r}_{ij})+V_{\rm OGE}(\vec{r}_{ij})+V_{\chi}(\vec{r}_{ij}) &  \\ 
\left[ q_iq_j=cn \right]  \Rightarrow V_{\rm CON}(\vec{r}_{ij})+V_{\rm OGE}(\vec{r}_{ij}) & 
\end{array} \right.\,,
\label{pot}
\end{equation}
where $n$ stands for the light quarks $u$ and $d$.
Notice that for the particular case of heavy quarks ($c$ or $b$) chiral symmetry is
explicitly broken and therefore boson exchanges do not contribute.
The parameters of the model are those of Ref.~\cite{Car12}.
The model guarantees a nice description of the baryon ($N$ and $\Delta$)~\cite{Valb05}
and the meson ($D$ and $D^*$) spectra~\cite{Vij05}.

In order to derive the local $B_n M_m\to B_k M_l$ interaction from the
basic $qq$ interaction defined above, we use a Born-Oppenheimer
approximation. Explicitly, the potential is calculated as follows,
\begin{equation}
V_{B_n M_m (L \, S \, T) \rightarrow B_k M_l (L^{\prime}\, S^{\prime}\, T)} (R) =
\xi_{L \,S \, T}^{L^{\prime}\, S^{\prime}\, T} (R) \, - \, \xi_{L \,S \,
T}^{L^{\prime}\, S^{\prime}\, T} (\infty) \, ,  \label{Poten1}
\end{equation}
\noindent where
\begin{equation}
\xi_{L \, S \, T}^{L^{\prime}\, S^{\prime}\, T} (R) \, = \, {\frac{{\left
\langle \Psi_{B_k M_l}^{L^{\prime}\, S^{\prime}\, T} ({\vec R}) \mid
\sum_{i<j=1}^{5} V_{q_iq_j}({\vec r}_{ij}) \mid \Psi_{B_n M_m}^{L \, S \, T} ({\vec R%
}) \right \rangle} }{{\sqrt{\left \langle \Psi_{B_k M_l }^{L^{\prime}\,
S^{\prime}\, T} ({\vec R}) \mid \Psi_{B_k M_l }^{L^{\prime}\, S^{\prime}\, T} ({%
\vec R}) \right \rangle} \sqrt{\left \langle \Psi_{B_n M_m }^{L \, S \, T} ({\vec %
R}) \mid \Psi_{B_n M_m }^{L \, S \, T} ({\vec R}) \right \rangle}}}} \, .
\label{Poten2}
\end{equation}
In the last expression the quark coordinates are integrated out keeping $R$
fixed, the resulting interaction being a function of the baryon-meson relative 
distance. The wave function $\Psi_{B_n M_m}^{L \, S \, T}({\vec R})$ for the baryon-meson
system has been discussed in Sec.~\ref{secIIA}.
\begin{figure}[b]
\vspace*{2cm}
\mbox{\epsfxsize=90mm\epsffile{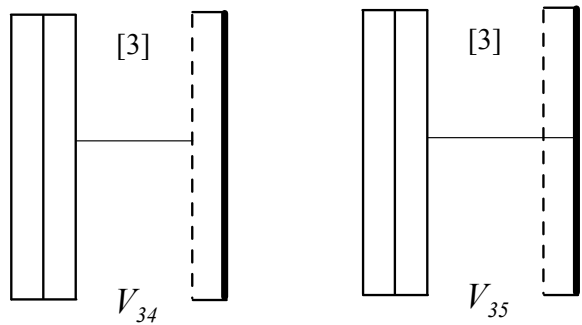}}
\vspace*{-4cm}
\caption{Different diagrams contributing to the
baryon-meson interaction. The vertical thin solid lines represent light quarks,
the vertical thick solid line represents a heavy quark, the vertical dashed line
stands for the light antiquark, and the horizontal solid line represents the exchanged particle.
The number between square brackets stands for
the number of diagrams topologically equivalent.}
\label{fig1}
\end{figure}

We show in Fig.~\ref{fig1} the different diagrams contributing to the baryon-meson interaction.
As compared to the $N\bar D$ case~\cite{Car12} and due to the absence of quark-exchange 
contributions, the number of diagrams is greatly reduced, getting just 
purely hadronic interactions.

\subsection{Integral equations for the two-body systems}
\label{secIV}
\begin{table}[t]
\caption{Interacting baryon-meson channels in the isospin-spin $(T,J$) basis. See text for details.}
\label{tab2}
\begin{tabular}{c|cccc}
\hline \hline
                         & $T=0$                                 &&  $T=1$                                & $T=2$          \\
\hline
\multirow{2}{*}{$J=1/2$} & $N D  - N D^*$                        && $N D - N D^* - \Delta D^* $           & $\Delta D^*$   \\
                         & $[\Sigma_c \pi -\Lambda^+_c \eta]$    && $[\Lambda^+_c \pi -\Sigma_c \pi]$     & $[\Sigma_c \pi]$ \\ \hline
\multirow{2}{*}{$J=3/2$} & $N D^* $                              && $N D^* - \Delta D - \Delta D^* $      & $\Delta D  - \Delta D^*$ \\
                         & $[\Sigma^*_c \pi -\Lambda^+_c \omega]$&& $[\Sigma^*_c \pi -\Lambda^+_c \rho]$  & $[\Sigma^*_c \pi]$ \\ \hline
\multirow{2}{*}{$J=5/2$} & $-$                                   && $\Delta  D^* $                        & $\Delta D^* $ \\ 
                         & $[\Sigma^*_c \rho]$                   && $[\Sigma^*_c \rho -\Sigma^*_c \omega]$& $[\Sigma^*_c \rho]$ \\
\hline\hline
\end{tabular}
\end{table}

\begin{figure}[b]
\vspace*{2.4cm}
\hspace*{5cm}\mbox{\epsfxsize=90mm\epsffile{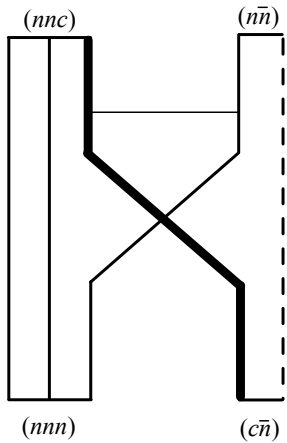}}
\vspace*{-3cm}
\caption{Diagram representing the coupling between a light baryon--charm meson channel
and a charm baryon--light meson two--hadron system.}
\label{fig2}
\end{figure}

To study the possible existence of molecular states made of a light baryon, $N$ or $\Delta$,
and a charmed meson, $D$ or $D^*$, we have solved the Lippmann-Schwinger equation 
for negative energies looking at the Fredholm determinant $D_F(E)$ at zero 
energy~\cite{Gar87}. If there are no interactions then $D_F(0)=1$, 
if the system is attractive then $D_F(0)<1$, and if a
bound state exists then $D_F(0)<0$. This method permitted 
us to obtain robust predictions even for zero-energy bound states, and gave
information about attractive channels that may lodge a resonance in similar systems~\cite{Car09}.
We consider a baryon-meson system $B_i M_j$ ($B_i=N$ or $\Delta$ and $M_j=D$ 
or $D^*$) in a relative $S$ state interacting through a potential $V$ that contains a
tensor force. Then, in general, there is a coupling to the 
$B_i M_j$ $D$ wave. Moreover, the baryon-meson system can couple to other 
baryon-meson states. We show in the first row of each spin cell of Table~\ref{tab2} the lowest light baryon--charm meson
coupled channels in the isospin-spin $(T,J)$ basis. They would contribute to our first approach, a coupled-channel
calculation of the $ND$ system looking for molecular states. As we have done
in Ref.~\cite{Car09}, we will later on allow for the rearrangement of quarks at short
distances giving rise to a coupling to a charm baryon--light meson two--hadron system,
through the diagram represented in Fig.~\ref{fig2}. For this case we show in the second row
of each spin cell of Table~\ref{tab2}, between square brackets, the adittional channels 
contributing to each $(T,J)$ state. They would contribute to the second calculation, looking
for compact five--quark states.

Thus, if we denote the different baryon-meson systems as channel $A_i$,
the Lippmann-Schwinger equation for the baryon-meson scattering becomes
\begin{eqnarray}
t_{\alpha\beta;TJ}^{\ell_\alpha s_\alpha, \ell_\beta s_\beta}(p_\alpha,p_\beta;E)& = & 
V_{\alpha\beta;TJ}^{\ell_\alpha s_\alpha, \ell_\beta s_\beta}(p_\alpha,p_\beta)+
\sum_{\gamma=A_1,A_2,\cdots}\sum_{\ell_\gamma=0,2} 
\int_0^\infty p_\gamma^2 dp_\gamma V_{\alpha\gamma;TJ}^{\ell_\alpha s_\alpha, \ell_\gamma s_\gamma}
(p_\alpha,p_\gamma) \nonumber \\
& \times& \, G_\gamma(E;p_\gamma)
t_{\gamma\beta;TJ}^{\ell_\gamma s_\gamma, \ell_\beta s_\beta}
(p_\gamma,p_\beta;E) \,\,\,\, , \, \alpha,\beta=A_1,A_2,\cdots \,\, ,
\label{eq0}
\end{eqnarray}
where $t$ is the two-body scattering amplitude, $T$, $J$, and $E$ are the
isospin, total angular momentum and energy of the system,
$\ell_{\alpha} s_{\alpha}$, $\ell_{\gamma} s_{\gamma}$, and
$\ell_{\beta} s_{\beta }$
are the initial, intermediate, and final orbital angular momentum and spin, respectively,
 and $p_\gamma$ is the relative momentum of the
two-body system $\gamma$. The propagators $G_\gamma(E;p_\gamma)$ are given by
\begin{equation}
G_\gamma(E;p_\gamma)=\frac{2 \mu_\gamma}{k^2_\gamma-p^2_\gamma + i \epsilon} \, ,
\end{equation}
with
\begin{equation}
E=\frac{k^2_\gamma}{2 \mu_\gamma} \, ,
\end{equation}
where $\mu_\gamma$ is the reduced mass of the two-body system $\gamma$.
For bound-state problems $E < 0$ so that the singularity of the propagator
is never touched and we can forget the $i\epsilon$ in the denominator.
If we make the change of variables
\begin{equation}
p_\gamma = d\frac{1+x_\gamma}{1-x_\gamma},
\label{eq2}
\end{equation}
where $d$ is a scale parameter, and the same for $p_\alpha$ and $p_\beta$, we can
write Eq.~(\ref{eq0}) as
\begin{eqnarray}
t_{\alpha\beta;TJ}^{\ell_\alpha s_\alpha, \ell_\beta s_\beta}(x_\alpha,x_\beta;E)& = & 
V_{\alpha\beta;TJ}^{\ell_\alpha s_\alpha, \ell_\beta s_\beta}(x_\alpha,x_\beta)+
\sum_{\gamma=A_1,A_2,\cdots}\sum_{\ell_\gamma=0,2} 
\int_{-1}^1 d^2\left(\frac{1+x_\gamma}{1-x_\gamma} \right)^2 \,\, \frac{2d}{(1-x_\gamma)^2} \,
dx_\gamma \nonumber \\
&\times & V_{\alpha\gamma;TJ}^{\ell_\alpha s_\alpha, \ell_\gamma s_\gamma}
(x_\alpha,x_\gamma) \, G_\gamma(E;p_\gamma) \,
t_{\gamma\beta;TJ}^{\ell_\gamma s_\gamma, \ell_\beta s_\beta}
(x_\gamma,x_\beta;E) \, .
\label{eq3}
\end{eqnarray}
We solve this equation by replacing the integral from $-1$ to $1$ by a
Gauss-Legendre quadrature which results in the set of
linear equations
\begin{equation}
\sum_{\gamma=A_1,A_2,\cdots}\sum_{\ell_\gamma=0,2}\sum_{m=1}^N
M_{\alpha\gamma;TJ}^{n \ell_\alpha s_\alpha, m \ell_\gamma s_\gamma}(E) \, 
t_{\gamma\beta;TJ}^{\ell_\gamma s_\gamma, \ell_\beta s_\beta}(x_m,x_k;E) =  
V_{\alpha\beta;TJ}^{\ell_\alpha s_\alpha, \ell_\beta s_\beta}(x_n,x_k) \, ,
\label{eq4}
\end{equation}
with
\begin{eqnarray}
M_{\alpha\gamma;TJ}^{n \ell_\alpha s_\alpha, m \ell_\gamma s_\gamma}(E)
& = & \delta_{nm}\delta_{\ell_\alpha \ell_\gamma} \delta_{s_\alpha s_\gamma}
- w_m d^2\left(\frac{1+x_m}{1-x_m}\right)^2 \frac{2d}{(1-x_m)^2} \nonumber \\
& \times & V_{\alpha\gamma;TJ}^{\ell_\alpha s_\alpha, \ell_\gamma s_\gamma}(x_n,x_m) 
\, G_\gamma(E;{p_\gamma}_m),
\label{eq5}
\end{eqnarray}
and where $w_m$ and $x_m$ are the weights and abscissas of the Gauss-Legendre
quadrature while ${p_\gamma}_m$ is obtained by putting
$x_\gamma=x_m$ in Eq.~(\ref{eq2}).
If a bound state exists at an energy $E_B$, the determinant of the matrix
$M_{\alpha\gamma;TJ}^{n \ell_\alpha s_\alpha, m \ell_\gamma s_\gamma}(E_B)$ 
vanishes, i.e., $\left|M_{\alpha\gamma;TJ}(E_B)\right|=0$.

\section{Results and discussion}
\label{secV}

Regarding the $ND$ interaction there are general trends that can be briefly summarized.
It is worth noting the absence of quark-exchange diagrams, that also prohibits the OGE contribution,
and thus quark-exchange effects are not present. Thus, the interaction 
comes determined by the OPE and OSE.
For very-long distances ($R>$ 4 fm) the dominant term is
the OPE potential, since it corresponds to the longest-range piece.
The OPE is also responsible altogether with the OSE for the
long-range part behavior (1.5 fm $<R<$ 4 fm), due to the combined effect of
shorter range and a bigger strength for the OSE as compared to the OPE.
The OSE gives the dominant contribution in the intermediate range (0.8 fm $<R<$ 1.5
fm), determining the attractive character of the potential in this region.
The short-range ($R<$ 0.8 fm) potential is either repulsive or attractive depending on the balance
between the OSE and OPE. Due to the nonexistence of quark-Pauli
correlations from the norm as well as from the interacting potential, one gets a
genuine baryonic interaction. Thus, dynamical quark-exchange effects do not play 
a relevant role in the $ND$ interaction unlike the $N \bar D$ case. 
\begin{table}[b]
\caption{Character of the interaction in the different $ND$ $(T,J)$ channels.}
\label{tab3}
\begin{tabular}{cccc}
\hline
\hline
        & $T=0$                &  $T=1$             & $T=2$ \\
\hline
$J=1/2$ &  Attractive          &  Weak              &  Weakly attractive   \\
$J=3/2$ &  Weak                &  Attractive        &  Attractive          \\
$J=5/2$ &  $-$                 &  Weakly attractive &  Attractive          \\ \hline \hline
\end{tabular}
\end{table}

\begin{figure}[t]
\vspace*{-7cm}
\hspace*{-1cm}\mbox{\epsfxsize=180mm\epsffile{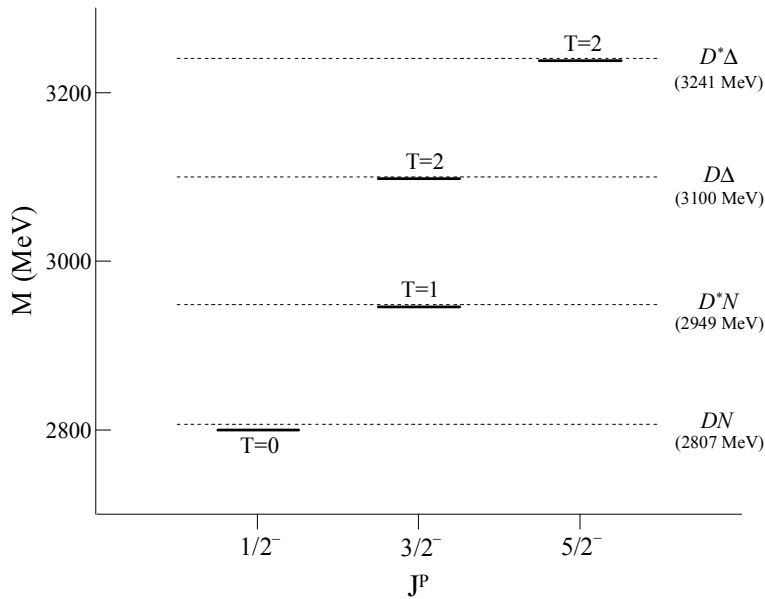}}
\vspace*{-12.cm}
\caption{Masses and quantum numbers of molecular $DN$ states. The dashed 
lines stand for the different two-hadron thresholds.}
\label{fig3}
\end{figure}

Using the interactions described above, we have solved the coupled-channel
problem of the baryon-meson systems made of a baryon, $N$ or $\Delta$, and a
meson, $D$ or $D^*$ as explained in Sec.~\ref{secIV}. The existence of 
bound states or resonances will generate baryonic states with charm $+1$ 
that could be identified as some of the excited states measured in the 
charm baryon spectrum. In Table~\ref{tab3} we summarize 
the character of the interaction in the different $(T,J)$ channels.
It can be observed that due to the absence of quark-Pauli correlations and the contribution
of the OSE the interaction is
in general attractive, giving rise to states that appear close to different
thresholds. We have represented in Fig.~\ref{fig3} the masses and quantum numbers
of the possible molecular $ND$ states.
The strongest interaction is obtained in the $ND^*$ $(T)J^P=(0)1/2^-$ channel, that it is 
coupled to the $ND$ $(T)J^P=(0)1/2^-$ partial wave (see Table~\ref{tab2}), generating
the best candidate to lodge a molecule. The expectation value of the isospin operator, $-3$ for isosinglet and $+1$
for isotriplet states, would reduce the attraction of isotriplet channels
as compared to attractive isosinglet channels with the same spin $J$ and vice versa, as can be easily
checked in the first two columns of Table~\ref{tab3}.

\begin{figure}[b]
%\vspace*{-7cm}
\mbox{\epsfxsize=80mm\epsffile{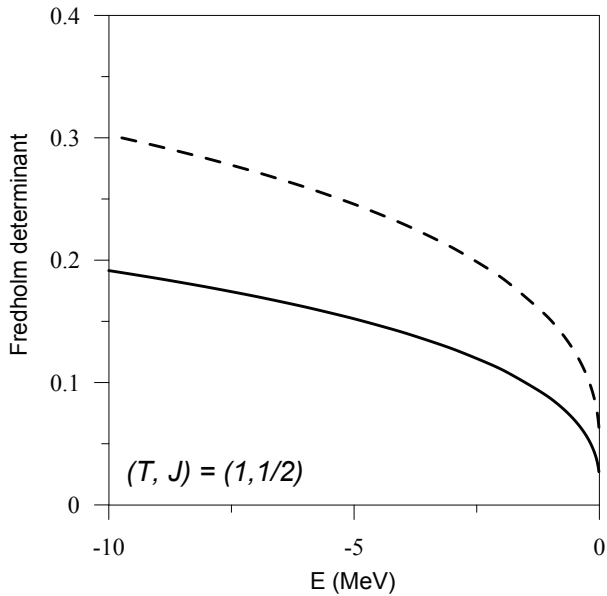}}
%\vspace*{-1.5cm}
\caption{Fredholm determinant of the $(T)J^P=(1)1/2^-$ channel considering 
all $ND$ contributions of Table~\ref{tab2} (solid line) and neglecting the $\Delta$
degrees of freedom (dashed line).}
\label{fig4}
\end{figure}

Our results may be compared to those of Ref.~\cite{Yam13} where the $N D$ system 
has been analyzed by means of a hadronic model using Lagrangians satisfying heavy 
quark symmetry and chiral symmetry. They arrive to the same
conclusion that the $(T)J^P=(0)1/2^-$ channel is the most attractive one. 
In Ref.~\cite{Yam13} this channel presents a bound
state of around 14.4 MeV for the model including only pion exchanges. 
The main difference of our results with those of Ref.~\cite{Yam13}
stem from the contribution of the scalar interaction and the consideration 
of explicit $\Delta$ degrees of freedom in our calculation. 
In a hadronic theory without explicit $\Delta$ degrees of freedom only a few channels 
survive and the coupled-channel dynamics would become simpler (see Table~\ref{tab2}). As a consequence,
for example, one could not get $T=2$ channels. In the chiral constituent quark model the
importance of the $\Delta$ degrees of freedom is known since long time ago~\cite{Val95,Nak98}.
It provides us with an isospin dependent mechanism that allows to correctly describe the
low-energy $NN$ $S$ wave phase-shifts through a coupled-channel effect, giving an
important attractive contribution for the $^1S_0$ $NN$ partial wave.
To emphasize the importance of coupled-channel dynamics~\cite{Lut05}, we have repeated 
the calculation explained in Sec.~\ref{secIV} for the $(T)J^P=(1)1/2^-$ channel but suppressing 
the states containing $\Delta's$. As can be seen in Fig.~\ref{fig4}, neglecting the $\Delta$ degrees of freedom the Fredholm
determinant gets larger, indicating a loss of attraction. For the single channel calculation the $(T)J^P=(1)1/2^-$
bound state does not appear, in agreement with the conclusions of Ref.~\cite{Yam13}. 

The $DN$ molecular states appearing in Fig.~\ref{fig3} could be an important ingredient of the charm baryon spectrum.
It has been recently suggested~\cite{Zha14} the possibility of the $\Sigma_c(2800)$ being an $S$ wave
$DN$ molecular state with $J^P=1/2^-$ and the $\Lambda_c(2940)^+$ an $S$ wave $D^*N$ state
with $J^P=3/2^-$, what would agree rather well the picture shown in Fig.~\ref{fig3}.
One may also find an experimental candidate for the $(T)J^P=(2)5/2^-$ resonance in one of the 
states recently reported by the BABAR Collaboration~\cite{Lee12}, an unexplained 
structure with a mass of 3250 MeV/c$^2$ in the $\Sigma_c^{++} \pi^- \pi^-$ invariant
mass. This state has also been recently suggested as a possible pentaquark~\cite{Alb13},
something that would be relevant in the second part of our discussion.
In spite of this agreement, one should note that the assignment of quantum numbers to baryon resonances on the 
charm baryon spectrum~\cite{ChaXX} and the identification of their internal 
structure~\cite{Hel07,Hel10,Don10,Yam13,Alb13,Zha14,ChaYY,Lut05} is still an open issue that
needs of further experimental analysis and also theoretical efforts. Such uncertainty 
has been recently revitalized by emphasizing the potential importance
of the relativistic kinematics of the light quark pair~\cite{Val14} casting doubts
even on the assignment of quantum numbers to experimental states just based on the 
non-relativistic quark-model.

\begin{figure}[t]
%\vspace*{-1cm}
\mbox{\epsfxsize=140mm\epsffile{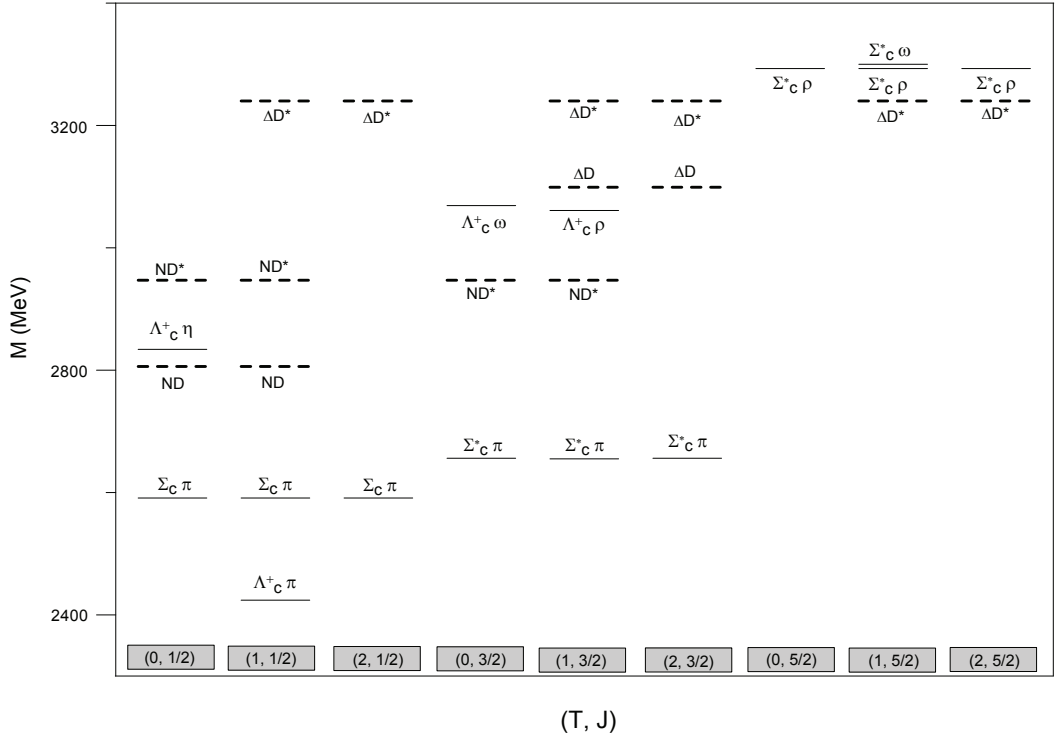}}
\vspace*{-0cm}
\caption{Different two-body channels contributing to each set of $(T,J)$ quantum numbers
as shown in Table~\protect\ref{tab2}.}
\label{fig5}
\end{figure}

One could also find contributions to the charm baryon spectrum with a more involved structure
such as compact five--quark states beyond simple $ND$ resonances~\cite{Vij09}.
The study of these contributions requires from a full coupled-channel approach including all
possible physical states contributing to a given set of quantum numbers $(T,J)$, as has
been demostrated in Ref.~\cite{Car11} for the charmonium spectrum.
Standard mesons ($q\bar q$) and baryons ($qqq$) are the only clusters of quarks where it is not possible to construct a 
color singlet using a subset of their constituents. Thus, $q\bar q$ and $qqq$ states are proper solutions 
of the two- and three-quark hamiltonian, respectively, corresponding in all cases to bound states. 
This, however, is not the case for multiquark combinations, and in particular for five--quark states
addressing the baryon spectrum. Thus, when dealing with higher order Fock space contributions to
baryon spectroscopy, one has to discriminate between possible five--quark bound states or resonances 
and simple pieces of the baryon--meson continuum. For this purpose, one has to analyze the two--hadron 
states that constitute the possible thresholds for each set of quantum numbers.
These thresholds have to be determined assuming quantum number conservation within exactly the same 
scheme (parameters and interactions) used for the five--body calculation. Working with strongly 
interacting particles, a baryon--meson state should have well--defined total angular momentum ($J$) 
and parity ($P$). If noncentral forces are not considered, orbital angular 
momentum ($L$) and total spin ($S$) are also good quantum numbers. We have represented in Fig.~\ref{fig5}
the different two-hadron thresholds contributing to each set of $(T,J)$ quantum numbers.

Given a general five--quark state contributing to the $ND$ wave function, ($nnnQ\bar n$)(in the following $n$ 
stands for a light quark and $Q$ for a heavy $c$ or $b$ quark), two different thresholds are 
allowed, $(nnn)(Q\bar n)$ and $(nnQ)(n\bar n)$.
A very simple property~\cite{Ber80} of the ground state solutions of the Schr\"odinger ($q_1\bar q_2$) two--body problem 
is that they are concave in $(m_{q_1}^{-1}+m_{q_2}^{-1})$, and hence $M_{Q\bar n}+M_{\bar Q n}\geqslant M_{Q\bar Q}+M_{n\bar n}$. 
This property is enforced both by nature\footnote{$M_{D^*}+M_{\bar D^*}=4014$ MeV $\geqslant M_{J/\psi}+M_{\omega}=3879$ MeV} 
and by all models in the literature unless forced to do otherwise. It implies that in all relevant 
cases the lowest two-meson threshold for any $(Q n \bar Q \bar n)$ state will be the one made of quarkonium-light 
mesons, i.e., $(Q\bar Q)(n\bar n)$ (see Fig. 1 of Ref.~\cite{Car12b}).
A straightforward generalization of this property to the five--quark system could be obtained within
a quark-diquark model if $m_{q_1} \le m_{q_2} \le m_{q_3}$. Then
$M_{q_3 \bar q_2} + M_{q_1 \bar q_1} \le M_{q_3 \bar q_1} + M_{q_1 \bar q_2}$,
because the intervals in $1/ \mu$ of the left hand side and right hand side
have the same middle, but the left hand side one is wider that the right hand side one.
Now, in a crude quark-diquark model, one can translate this as
$M_{q_3 q_1 q_1} + M_{q_1 \bar q_1} \le M_{q_3 \bar q_1} + M_{q_1 q_1 q_1}$, as it is 
observed in Fig.~\ref{fig5} except for the higher spin states where the angular momentum 
coupling rules impose further restrictions.~\footnote{We thank
to J.~M. Richard for this simple and nice argument that does not make any assumption on the shape of 
the interaction, linear or not, although it assumes a quark-diquark ansatz.}
\begin{table}[b]
\caption{Character of the interaction in the different baryon-meson $(T,J)$ channels.}
\label{tab4}
\begin{tabular}{cccc}
\hline
\hline
        & $T=0$                &  $T=1$             & $T=2$ \\
\hline
$J=1/2$ &  Weakly attractive       &  Weak              &  Strongly repulsive      \\
$J=3/2$ &  Weak                    &  Weak              &  Weak                    \\
$J=5/2$ &  Attractive              &  Attractive        &  Attractive              \\ \hline \hline
\end{tabular}
\end{table}
\begin{figure}[t]
\vspace*{-7cm}
\hspace*{-1cm}\mbox{\epsfxsize=180mm\epsffile{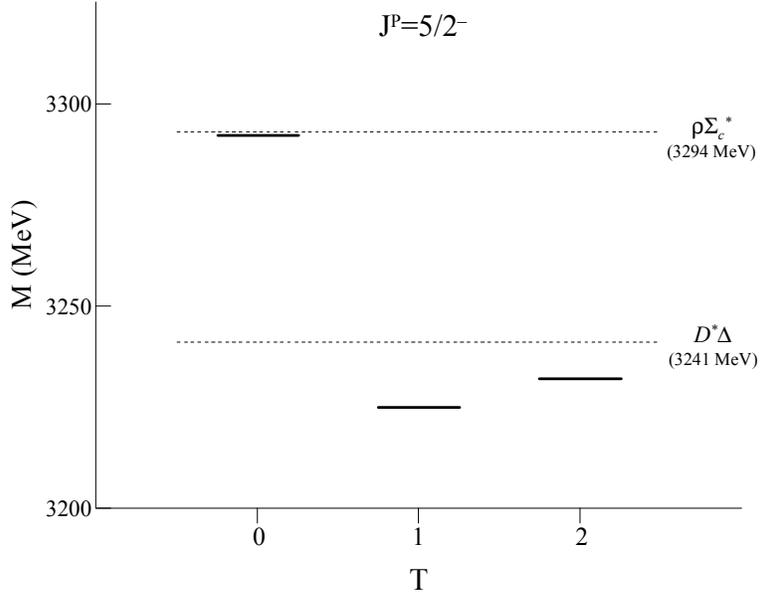}}
\vspace*{-12.cm}
\caption{Masses and quantum numbers of compact five--quark states. The dashed 
lines stand for the different two-hadron thresholds.}
\label{fig6}
\end{figure}
\begin{figure}[b]
%\vspace*{-7cm}
\mbox{\epsfxsize=80mm\epsffile{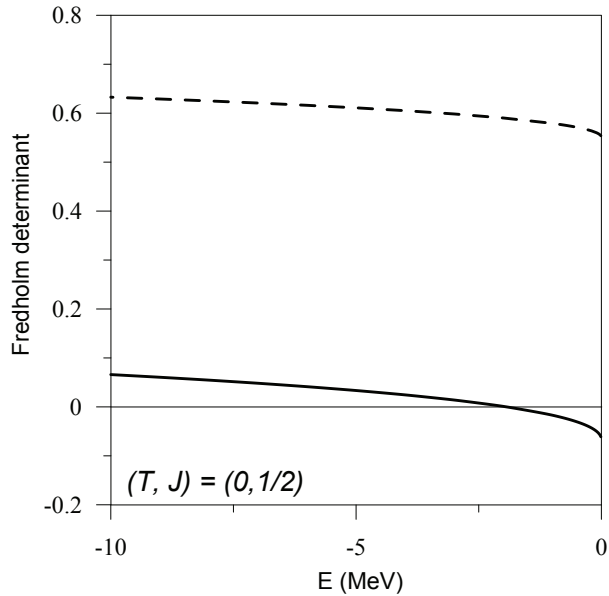}}
%\vspace*{-1.5cm}
\caption{Fredholm determinant of the $(T)J^P=(0)1/2^-$ channel considering 
only the $ND$ contributions of Table~\ref{tab2} (solid line) and considering all
contributions (dashed line). $E$ indicates the energy below the corresponding
lowest threshold, $ND$ for the solid line and $\Sigma_c\pi$ for the dashed one,
as can be seen in Fig.~\ref{fig5}.}
\label{fig7}
\end{figure}

Hence, as we have already illustrated in Fig.~\ref{fig4}, an important source of attraction might be
the coupled-channel effect of the two thresholds, $(nnn)(Q\bar n)\leftrightarrow(nnQ)(n\bar n)$~\cite{Lut05}.
Thus, to check the efficiency of this mechanism, we have repeated the calculation
of Sec.~\ref{secIV} but considering all physical states reflected in Table~\ref{tab2}.
We have represented in Fig.~\ref{fig5} the lowest baryon-meson thresholds contributing to each
set of $(T,J)$ quantum numbers. In Table~\ref{tab4} we have summarized the character 
of the interaction in the different $(T,J)$ channels.
When the $(nnn)(Q\bar n)$ and $(nnQ)(n\bar n)$ thresholds are sufficiently 
far away, the coupled-channel effect is small, and bound states are not found. 
However, when the thresholds move closer, the coupled-channel strength is increased, 
and bound states may appear for a subset of 
quantum numbers. Hence, threshold vicinity is a required but not sufficient 
condition to bind a five--quark state. Under these conditions, there are the channels 
with high spin $J^P=5/2^-$ the only ones that may lodge a 
compact five-quark state for all isospins as it is shown in Fig.~\ref{fig6}. The reason stems on the reverse of the ordering of the
thresholds, being the lowest threshold $(nnn)(Q\bar n)$ the one with the more attractive
interaction. In the other cases, the break apart
threshold $(nnQ)(n\bar n)$, weakly interacting, is the lowest one destroying the possibility for any resonance.
This is illustrated in Fig.~\ref{fig7} where we show the Fredholm determinant for the $(T)J^P=(0)1/2^-$ channel. 
When one only considers $ND$ channels (solid line) the interaction is attractive and it has a bound state.
However, when the lowest break apart threshold is considered (dashed line) the bound state does not
appear any more.

As already advertised in the first
part of our discussion, of particular interest is the $(T)J^P=(2)5/2^-$ state, that survives
the consideration of the break apart thresholds. It may
correspond to the $\Theta_c(3250)$ pentaquark found by the QCD sum rule analysis of Ref.~\cite{Alb13}
when studying the unexplained structure with a mass of 3250 MeV/c$^2$ in the $\Sigma_c^{++} \pi^- \pi^-$ 
invariant mass reported recently by the BABAR Collaboration~\cite{Lee12}. 
Such state could be also detected by the propagation of $D$ mesons in nuclear matter as
an $S$ wave $\Delta D^*$ system and it thus constitutes a challenge for the $\bar P$ANDA Collaboration.

\section{Summary}
\label{secVI}

Summarizing, we have studied higher order Fock space components on the charm baryon 
spectrum. For this purpose we have used two different approaches. In a first step
we did a coupled-channel calculation of the $ND$ system looking for molecular states. 
In a second step we allowed for the coupling to a heavy baryon--light meson system
looking for compact exotic five--quark states. Both calculations have been done
within the framework of a full-fledged chiral constituent quark model.
This model, tuned in the description of the baryon and meson spectra 
as well as the $NN$ interaction, provides parameter-free predictions for 
charm $+1$ molecular or compact two-hadron systems. 
Unlike the $N \bar D$ system, no sharp quark-Pauli effects are found
due to the non-existence of quark-exchange diagrams
and thus of the OGE contribution.
The importance of the coupled-channel dynamics has
been emphasized to connect with the result of other hadronic models.
We have found several close to threshold resonances in the $ND$ system that
could be traced back to some of the measured charm baryon excited states. If the
full dynamics of the five--quark system is considered the number of resonances is
reduced and their energies augmented. Of particular interest is the prediction
of a $(T)J^P=(2)5/2^-$ baryonic state, that survives
the consideration of the break apart thresholds. It may
correspond to the $\Theta_c(3250)$ pentaquark found by the QCD sum rule analysis of Ref.~\cite{Alb13}
when studying the unexplained structure with a mass of 3250 MeV/c$^2$ in the $\Sigma_c^{++} \pi^- \pi^-$ 
invariant mass reported recently by the BABAR Collaboration~\cite{Lee12}. 

The advent of new experimental data on charm baryon spectroscopy will shed light about
the structure of some of the already observed states that, otherwise, will also help us in
understanding the short-range dynamics of many--quarks systems (confinement), either confirming
the existence of $ND$ resonances close-to-threshold or not. The first scenario will point
to a two-hadron resonance while the second may be a hint for the presence of many--quark
states. This objective may be attainable in several current and future Collaborations like
$\bar {\rm P}$ANDA, LHCb, ExHIC or J-PARC.

\section{Acknowledgments}
The authors thank to Prof. J. Vijande and Prof. J.-M. Richard for discussions
on several issues that are addressed along this work.
This work has been partially funded by the Spanish Ministerio de
Educaci\'on y Ciencia and EU FEDER under Contract No. FPA2010-21750-C02-02,
and by the Spanish Consolider-Ingenio 2010 Program CPAN (CSD2007-00042).


\begin{thebibliography}{99}

\bibitem{Bra13} E.~Braaten,
			Phys. Rev. Lett. {\bf 111}, 162003 (2013).
			
\bibitem{Cho03} S.~-K.~Choi {\it et al.} (Belle Collaboration), 
			Phys. Rev. Lett. {\bf 91}, 262001 (2003).

\bibitem{Gel64} M.~Gell-Mann,
			Phys. Lett. {\bf 8}, 214 (1964). 
			
\bibitem{Bod13} G.~T.~Bodwin, E.~Braaten, E.~Eichten, S.~L.~Olsen, T.~K.~Pedlar, and J.~Russ,
		  arXiv:1307.7425, and references therein.

\bibitem{Kle10} E.~Klempt and J.~-M.~Richard,
			Rev. Mod. Phys. {\bf 82}, 1095 (2010). 

\bibitem{Cre13} V.~Crede and W.~Roberts, 
			Rept. Prog. Phys. {\bf 76}, 076301 (2013). 
		  
\bibitem{Hel07} X.~G.~He, X.~Q.~Li, X.~Liu, and X.~Q.~Zeng, 
			Eur. Phys. J. C {\bf 51}, 883 (2007).

\bibitem{Hel10} J.~He and X.~Liu, 
			Phys. Rev. D {\bf 82}, 114029 (2010).

\bibitem{Don10} Y.~Dong, A.~Faessler, T.~Gutsche, and V.~E.~Lyubovitskij,
			Phys. Rev. D {\bf 81}, 074011 (2010).

\bibitem{Yam13} Y.~Yamaguchi, S.~Ohkoda, S.~Yasui, and A.~Hosaka,
			Phys. Rev. D {\bf 87}, 074019 (2013).

\bibitem{Alb13} R.~M.~Albuquerque, S.~H.~Lee, and M.~Nielsen,
			Phys. Rev. D {\bf 88}, 076001 (2013).
						
\bibitem{Zha14} J.~-R.~Zhang,
			Phys. Rev. D {\bf 89}, 096006 (2014).
			
\bibitem{ChaYY}	C.~E.~Jim\'enez-Tejero, A.~Ramos, and I.~Vida\~na, 
			Phys. Rev. C {\bf 80}, 055206 (2009);	
								C.~Garc{\'\i}a-Recio, V.~K.~Magas, T.~Mizutani, J.~Nieves, A.~Ramos, L.~L.~Salcedo, and L.~Tolos, 
			Phys. Rev. D {\bf 79}, 054004 (2009);
			          O.~Romanets, L.~Tolos, C.~Garc{\'\i}a-Recio, J.~Nieves, L.~L.~Salcedo, and R.~G.~E.~Timmermans, 
			Phys. Rev. D {\bf 85}, 114032 (2012);			          
								M.~Bayar, C.~W.~Xiao, T.~Hyodo, A.~Dot\'e, M.~Oka, and E.~Oset, 
			Phys. Rev. C {\bf 86}, 044004 (2012);	
								P.~G.~Ortega, D.~R.~Entem, and F.~Fern\'andez,
			Phys. Lett. B {\bf 718}, 1381 (2013).

\bibitem{Gon08} P.~Gonz\'alez, J.~Vijande, and A.~Valcarce,
      Phys. Rev. C {\bf 77}, 065213 (2008).
      
\bibitem{Pan09} U.~Wiedner ({$\bar {\rm P}$}ANDA Collaboration), 
{\it Future Prospects for Hadron Physics at $\bar P$ANDA},
			Prog. Part. Nucl. Phys. {\bf 66}, 477 (2011).
			
\bibitem{Kog83} J.~Kogut, M.~Stone, H.~W.~Wyld, W.~R.~Gibbs, J.~Shigemitsu,
S.~H.~Shenker, and D.~K.~Sinclair,
			Phys. Rev. Lett. {\bf 50}, 393 (1983).

\bibitem{Mat86} T.~Matsui and H.~Satz,
			Phys. Lett. B {\bf 178}, 416 (1986). 

\bibitem{Dov77} C.~B.~Dover and S.~H.~Kahana, 
			Phys. Rev. Lett. {\bf 39}, 1506 (1977).
			
\bibitem{Yok14} A.~Yokota, E.~Hiyama, and M.~Oka,
			Few-Body Syst., {\it in press} (2014) [DOI 10.1007/s00601-014-0895-2].

\bibitem{Kho12} A.~Khodjamirian, Ch.~Klein, Th.~Mannel, and Y.-M.~Wang,
			Eur. Phys. J. A {\bf 48}, 31 (2012).

\bibitem{Heo11} J.~He, Z.~Ouyang, X.~Liu, and X.~-Q.~Li, 
			Phys. Rev. D {\bf 84}, 114010 (2011).

\bibitem{Hai11} J.~Haidenbauer and G.~Krein,
				Phys. Lett. B {\bf 687}, 314 (2010).
			
\bibitem{Cho11} S.~Cho {\it et al.} (ExHIC Collaboration), 
			Phys. Rev. Lett. {\bf 106}, 212001 (2011).
			
\bibitem{Cho11b} S.~Cho {\it et al.} (ExHIC Collaboration), 
			Phys. Rev. C {\bf 84}, 064910 (2011).		
			
\bibitem{Mil13} D.~Milanes (LHCb Collaboration),
			Nucl. Phys. Proc. Suppl. {\bf 234}, 139 (2013). 
			
\bibitem{Aal14} T.~A.~Aaltonen {\it et al.} (CDF Collaboration),			
			Phys. Rev. D {\bf 89}, 072014 (2014). 

\bibitem{Sai07} K.~Saito, K.~Tsushima, and A.~W.~Thomas,
			Prog. Part. Nucl. Phys. {\bf 58}, 1 (2007).
			
\bibitem{Car14} T.~F.~Caram\'es, C.~E.~Fontoura, K.~Tsushima, G.~Krein, and A.~Valcarce,
			in preparation.
				
\bibitem{Val05} A.~Valcarce, H.~Garcilazo, F.~Fern\'andez, and P.~Gonz\'alez,
			Rept. Prog. Phys. {\bf 68}, 965 (2005).

\bibitem{Vij05} J.~Vijande, F.~Fern\'andez, and A.~Valcarce,
			J. Phys. G {\bf 31}, 481 (2005). 

\bibitem{Car09} T.~Fern\'andez-Caram\'es, A.~Valcarce, and J.~Vijande,
			Phys. Rev. Lett. {\bf 103}, 222001 (2009).
			
\bibitem{Car12} T.~F.~Caram\'es and A.~Valcarce,
			Phys. Rev. D {\bf 85}, 094017 (2012).
				
\bibitem{Ruj75} A.~de~R\'ujula, H.~Georgi, and S.~L.~Glashow,
			Phys. Rev. D {\bf 12}, 147 (1975).

\bibitem{Bal01} G.~S.~Bali, 
			Phys. Rep. {\bf 343}, 1 (2001).

\bibitem{Valb05} A.~Valcarce, H.~Garcilazo, and J.~Vijande,
			Phys. Rev. C {\bf 72}, 025206 (2005).

\bibitem{Gar87} H.~Garcilazo, 
			J. Phys. G {\bf 13}, L63 (1987).
			
\bibitem{Val95} A.~Valcarce, A.~Faessler, and F.~Fern\'andez,
			Phys. Lett. B {\bf 345}, 367 (1995).
			
\bibitem{Nak98} C.~Nakamoto and H.~Toki,
			Prog. Theor. Phys. {\bf 99}, 1001 (1998). 

\bibitem{Lut05} M.~F.~M.~Lutz and E.~E.~Kolomeitsev,
			Nucl. Phys. A {\bf 755}, 29c (2005).
			
\bibitem{Lee12} J.~P.~Lees {\it et al.}, (BABAR Collaboration)
			Phys. Rev. D {\bf 86}, 091102(R) (2012).
			
\bibitem{ChaXX} S.~Capstick and N.~Isgur, 
			Phys. Rev. D {\bf 34}, 2809 (1986);
								D.~Ebert, R.~N.~Faustov, and V.~O.~Galkin, 
			Phys. Lett. B {\bf 659}, 612 (2008);
								A.~Valcarce, H.~Garcilazo, and J.~Vijande,
			Eur. Phys. J. A {\bf 37}, 217 (2008).

\bibitem{Val14} A.~Valcarce, H.~Garcilazo, and J.~Vijande,
			Phys. Lett. B {\bf 733}, 288 (2014).

\bibitem{Vij09} J.~Vijande and A.~Valcarce,
      Phys. Rev. C {\bf 80}, 035204 (2009).
      
\bibitem{Car11} T.~F.~Caram\'es, A.~Valcarce, and J.~Vijande,
			Phys. Lett. B {\bf 699}, 291 (2011).	  

\bibitem{Ber80} R.~A.~Bertlmann and A.~Martin,
	    Nucl. Phys. B {\bf 168}, 111 (1980);
	    				  S.~Nussinov,
	    Z. Phys. C {\bf 3}, 165 (1979). 
                      
 \bibitem{Car12b} T.~F.~Caram\'es, A.~Valcarce, and J.~Vijande,
			Phys. Lett. B {\bf 709}, 358 (2012).	                      
                      
\end{thebibliography}
\end{document}